\documentclass[12pt]{article}

\usepackage{geometry}
\usepackage{amsmath}
\usepackage{amsthm}
\usepackage{amsfonts,dsfont}
\usepackage{amssymb}
\usepackage{stmaryrd}
\usepackage[all,cmtip]{xy}
\usepackage{fancyhdr}
\usepackage{verbatim}
\usepackage{color}
 \usepackage{graphicx}
 \usepackage[lofdepth,lotdepth]{subfig}

\theoremstyle{definition}\newtheorem{definition}{Definition}[section]
\theoremstyle{definition}\newtheorem{proposition}[definition]{Proposition}
\theoremstyle{definition}
\theoremstyle{definition}\newtheorem{theorem}[definition]{Theorem}
\theoremstyle{definition}
\theoremstyle{definition}
\theoremstyle{definition}\newtheorem{example}[definition]{Example}

\numberwithin{equation}{section}

\newcommand{\pp}[2]{\frac{\partial #1}{\partial #2}}

\newcommand{\n}{^{(n)}}

\newcommand{\vv}{\mathbf{v}}

\newcommand{\J}{{\rm J}}
\newcommand{\Jn}{\J^n}

\newcommand{\jn}{{\rm j}_n}

\def\comp{\raise 1pt \hbox{$\,\scriptstyle\circ\,$}}

\begin{document}

%%%%%%%%%
% Opening page
%%%%%%%%% 

\thispagestyle{fancy}
\fancyhead{}
\fancyfoot{}
\renewcommand{\headrulewidth}{0pt}
\cfoot{\thepage}
\rfoot{\today}

\vskip 1cm
\begin{center}
{\Large Symmetry Preserving Numerical Schemes for Partial Differential Equations and their Numerical Tests}
\vskip 1cm

\begin{tabular*}{1.0\textwidth}{@{\extracolsep{\fill}} ll}
Rapha\"el Rebelo\footnotemark[1] & Francis Valiquette\footnotemark[2]\\
D\'epartement de math\'ematiques et & Department of Mathematics and Statistics\\
de statistique & Dalhousie University\\
Universit\'e de Montr\'eal & 
Halifax, Nova Scotia, Canada \quad B3H 3J5\\
Montr\'eal, Qu\'ebec, Canada \quad H3C 3J7 & {\tt francisv@mathstat.dal.ca}\\
{\tt raph.rebelo@gmail.com} & {\tt http://www.mathstat.dal.ca/$\sim$francisv} 
\end{tabular*}
\end{center}

\footnotetext[1]{{\it Supported by a FQRNT Doctoral Research Scholarship.}}
\footnotetext[2]{{\it Supported by a NSERC of Canada Postdoctoral Fellowship.}}

\vskip 0.5cm\noindent
{\bf Mathematics subject classification (MSC2010)}:  58J70, 68N06

\vskip 1cm

\abstract{}
The method of equivariant moving frames on multi-space is used to construct symmetry preserving finite difference schemes of partial differential equations invariant under finite-dimensional symmetry groups.  Invariant numerical schemes for a heat equation with a logarithmic source and the spherical Burgers equation are obtained.  Numerical tests show how invariant schemes can be more accurate than standard discretizations on uniform rectangular meshes.

%%%%%
\section{Introduction}
%%%%%

In modern numerical analysis, much effort has been invested into developing  geometric integrators that incorporate geometrical structures of the system of differential equations being approximated. Well known examples include symplectic integrators \cite{CS-1990}, energy preserving methods \cite{QM-2008},  Lie--Poisson preserving methods \cite{ZM-1988}, and symmetry preserving numerical schemes \cite{BRW-2008, LW-2006, VW-2005}.  The motivation behind all this work is that, as a rule of thumb, geometric integrators give better results than many other standard numerical methods since they take into account qualitative properties of the system being studied.

For ordinary differential equations, the problem of generating invariant numerical schemes preserving the point symmetries of the original equations is now well-understood \cite{BCW-2006, BRW-2008, KO-2004, RW-2009}. There exists essentially two different methods for generating invariant finite difference schemes of differential equations.  The first approach, mainly developed by Dorodnitsyn, Levi and Winternitz, is based on Lie's infinitesimal symmetry method \cite{BD-2001, D-1991, D-1994, D-2010, DKW-2000, LW-2006, VW-2005}.  This approach makes use of Lie's infinitesimal symmetry criterion \cite{O-1993} to obtain finite difference invariants from which an invariant scheme is constructed by finding a combination of the invariants that converges, in the continuous limit, to the original system of differential equations.  The second approach, developed by Olver and Kim, consists of using the method of equivariant moving frames  \cite{K-2007, K-2008, O-2001}.   With Lie's infinitesimal method, the construction of invariant schemes can sometimes require a lot of work and insights, on the other hand, with the moving frame method the construction is completely algorithmic.  In both cases, the methods have proven their efficiency for generating invariant numerical schemes for ordinary differential equations.  In comparison, fewer applications involving partial differential equations can be found in the literature \cite{BD-2001, D-1991, D-1994, K-2008, VW-2005}.

In this paper we use the equivariant moving frame method to generate invariant finite difference schemes for partial differential equations with finite-dimensional symmetry groups.  To implement the moving frame method, the first step is to obtain appropriate approximations of the partial derivatives on an arbitrary mesh.  In \cite{O-2006}, Olver proposes a new approach to the theory of interpolation of functions of several variables, based on non-commutative quasi-determinants, to obtain these approximations.  We show here that one can use standard Taylor polynomial approximations to achieve the same goal in a somewhat simpler way.   The formulas obtained suggest an interesting interpretation of the continuous limit of the finite difference derivatives.  More details are given in Section \ref{discrete derivatives}.

Since the proper geometrical space for the numerical analysis of differential equations is that of multi-space \cite{O-2001}, we begin with a review of multi-space in Section \ref{Differential Equations and Numerical Schemes}.  The formulas used to approximate derivatives on an arbitrary mesh are introduced in Section \ref{discrete derivatives}. With this in hand, we review the multi-moving frame construction in Section \ref{mf section}.  As with the standard moving frame method, the equivariant multi-frame construction relies on Cartan's normalization of the group parameters of the action.  Given a multi-frame, there is a canonical invariantization map which projects finite difference expressions onto their invariant finite difference counterpart.  In particular, the invariantization of the finite difference derivatives gives finite difference invariants which, in the continuous limit, converge to the normalized differential invariants \cite{FO-1999}.  Thus, by rewriting a system of partial differential equations in terms of the normalized differential invariants, an invariant numerical scheme is obtained by replacing the normalized differential invariants by their invariant finite difference approximations.  The method is illustrated in Section 5 where invariant numerical schemes are obtained for a heat equation with a logarithmic source and for the spherical Burgers' equation. Section 6 is dedicated to numerical tests in which the precision of completely invariant schemes is compared to invariant and standard discretizations on a uniform mesh.

%%%%%
\section{Differential Equations and Numerical Schemes}\label{Differential Equations and Numerical Schemes}
%%%%%

Let $M$ be an $m$-dimensional manifold.  For $0 \leq n \leq \infty$ let $\Jn=\Jn(M,p)$ denote the \emph{extended $n^{\text{th}}$ order jet space} of $1\leq p < m$ dimensional submanifolds $S\subset M$.  The jet space is defined as the space of equivalence classes of submanifolds under the equivalence relation of $n^{\text{th}}$ contact at a point \cite{O-1993}.  We use the notation $\jn S|_z$ to denote the $n$-jet of $S$ at $z\in S$.  Given a submanifold $S \subset M$, we introduce the local coordinates $z=(x,u)=(x^1,\ldots,x^p,u^1,\ldots,u^q)$, with $q=m-p$, so that $S$ is locally the graph of a function $f(x)$:  $S=\{ (x,f(x))\}$.    In this coordinate chart the coordinate jets of $\jn S$ are $z\n=(x,u\n)$, where $u\n$ denotes the collection of derivatives $u^\alpha_J=\partial^k u^\alpha/ (\partial x)^J$ with $0\leq k=\#J \leq n$.

The approximation of the $n$-jet of a submanifold by finite difference derivatives is based on the evaluation of the submanifold at several points.  For reasons which will become clearer in the next section, we label the sample points with a multi-index $z_{N}=(x_N,u_N)=(x_N,f(x_N))$ where $N=(n^1,\ldots,n^p) \in \mathbb{Z}^p$.  For the finite difference derivatives to be well defined, the sample points must be distinct in the independent variables.  Thus we introduce the $k$-fold \emph{joint product}
$$
M^{\diamond k}=\{ (z_{N_1},\ldots,z_{N_k})| \, x_{N_i}\neq x_{N_j}\, \text{for all } i\neq j\} 
$$
which is the subset of the $k$-fold Cartesian product $M^{\times k}$. 

A smooth function $\Delta\colon \Jn \to \mathbb{R}$ on (an open subset of) the jet space is known as a  \emph{differential function}.  A system of differential equations is defined by the vanishing of one or more differential functions.  In the local coordinates $z\n=(x,u\n)$  a system of differential equations is given by
\begin{equation}\label{pde1}
\Delta_1(x,u\n)= \cdots =\Delta_\ell(x,u\n)=0.
\end{equation}

\begin{definition}\label{numerical scheme definition}
Let $\Delta_\nu(x,u\n)=0$, $\nu=1,\ldots, \ell$, be a system of differential equations.  A \emph{finite difference numerical scheme} is a system of equations
\begin{equation}\label{numerical scheme}
E_\mu(z_{N_1},\ldots,z_{N_k})=0,\qquad \mu=1,\ldots,\ell,\ldots, \ell+l.\nonumber
\end{equation}
defined on the joint product $M^{\diamond k}$ with the property that in the coalescent limit (continuous limit) $z_{N_i} \to z$, $E_{\mu}(z_{N_1},\ldots,z_{N_k}) \to \Delta_\nu(x,u\n)$ for $\mu=1,\ldots,\ell$ and $E_\mu \to 0$ for $\mu=\ell+1,\ldots, \ell+l$. The last $l$ equations $E_{\ell+1}=0,\ldots,E_{\ell+l}=0$ impose constraints on the mesh.
\end{definition}

There are two restrictions on the equations $E_{\ell+1}=0,\ldots,E_{\ell+l}=0$.  Firstly, the equations must be compatible so that, provided appropriate initial conditions, the independent variables $x_{N_i}$ are uniquely defined.  Secondly, these equations should not impose any restriction on the dependent variables $u_{N_i}$.  When these two conditions are satisfied we will say that the mesh equations are \emph{compatible}.   

In Definition \ref{numerical scheme definition}, the number of copies $k$ in the joint product will depend on the order of approximation of the numerical scheme.  The minimal number of sample points required to approximate the $n^\text{th}$ order system of equations \eqref{pde1} is $k=q\binom{p+n}{n}=\text{dim }\Jn - p$.  More sample points can be added for better numerical results . 

\begin{example}\label{standard heat}
A standard numerical scheme for the heat equation 
\begin{equation}\label{heat}
\Delta(x,t,u^{(2)}) = u_t  - u_{xx} - u \ln u=0
\end{equation}
with logarithmic source on the uniform rectangular mesh
\begin{equation}\label{uniform rectangular mesh}
x_{m,n}=h\, m + x_0,\qquad t_{m,n}= k \, n + t_0,\qquad \text{where} \qquad h, k, x_0, t_0\qquad \text{are constants}
\end{equation}
and $m$, $n$ are integers is given by 
\begin{subequations}
\begin{equation}\label{heat numerical scheme}
E_1=\Delta_t u - \Delta_{x}^2u - u_{m,n} \ln u_{m,n}=0,
\end{equation}
where
$$
\Delta_t u=\frac{u_{m,n+1}-u_{m,n}}{k},\qquad
\Delta_{x}^2u= \frac{u_{m+2,n}-2u_{m+1,n}+u_{m,n}}{h^2}
$$
are the standard finite difference derivatives on the uniform rectangular mesh \eqref{uniform rectangular mesh}.   In terms of Definition \ref{numerical scheme definition}, the mesh \eqref{uniform rectangular mesh} is defined by the equations
\begin{equation}\label{heat rectangular mesh}
E_{2,3,4,5}=\left\{ 
\begin{aligned}
&x_{m+1,n}-x_{m,n}=h, & \qquad & x_{m,n+1}-x_{m,n}=0,\\
&t_{m+1,n}-t_{m,n}=0, & & t_{m,n+1}-t_{m,n}=k.
\end{aligned}
\right.
\end{equation}
\end{subequations}
Given the initial conditions $x_0$, $t_0$ the equations \eqref{heat rectangular mesh} uniquely specify the points $(x_{m,n},t_{m,n})$.
\end{example}

Now, let $G$ be an $r$-dimensional Lie group acting regularly on $M$.  Throughout the paper, we will consistently use lower case letters to denote the source coordinates of the action, and capital letters to denote the target coordinates:
\begin{equation}\label{group action}
X^i=g \cdot x^i,\qquad U^\alpha= g \cdot u^\alpha,\qquad i=1,\ldots,p, \quad \alpha=1,\ldots,q, \qquad g\in G.
\end{equation}
Since a Lie group action preserves the contact equivalence of submanifolds, it induces an action on $\Jn$:
\begin{equation}\label{prolonged action}
g\cdot \jn S|_z = \jn (g\cdot S)|_{g\cdot z},\qquad g\in G,
\end{equation}
called the $n^\text{th}$ order prolonged action.  In applications, the prolonged action \eqref{prolonged action} is obtained by implementing the chain rule.  In local coordinates, we use the notation $(X,U\n)=g\cdot(x,u\n)$ to denote the prolonged action.  The Lie group $G$ also induces a natural action on the $k$-fold Cartesian product $M^{\times k}$ given by the product action $G^{\times k}$:
\begin{equation}\label{product action}
g\cdot (z_{N_1},\ldots, z_{N_k})=(g\cdot z_{N_1},\ldots, g\cdot z_{N_k}).
\end{equation}

\begin{definition}\label{symmetry definition}
Let $G$ be a Lie group acting regularly on $M\simeq X\times U$ and $\Delta_\nu(x,u\n)=0$ a system of differential equations.  The system of differential equations is said to be \emph{$G$-invariant} if $\Delta_\nu(g\cdot(x,u\n))=\Delta_\nu(x,u\n)$ for all $g \in G$.  Similarly, let $E_\mu(z_{N_1},\ldots,z_{N_k})=0$ be a system of finite difference equations.  Then it is $G$-invariant if $E_\mu(g\cdot(z_{N_1},$ $\ldots,$ $z_{N_k}))=E_\mu(z_{N_1},\ldots,z_{N_k})$ for all $g \in G$.
\end{definition}

%\begin{definition}\label{symmetry definition}
%Let $G$ be a Lie group acting regularly on $M$ and $\Delta_\nu(x,u\n)=0$ a system of differential equations.  Then $G$ is a symmetry group of the system of differential equations if for all $g\in G$ we have $\Delta_\nu(X,U\n)=0$ whenever $\Delta_\nu(x,u\n)=0$.  Similarly, let $E_\mu(z_{N_1},\ldots,z_{N_k})=0$ be a numerical scheme.  Then $G$ is a symmetry group of the numerical scheme if for all $g\in G$ we have $E_\mu(Z_{N_1},\ldots,Z_{N_k})=0$ whenever $E_\mu(z_{N_1},\ldots,z_{N_k})=0$.
%\end{definition}

%\begin{remark}
%Definition \ref{symmetry definition} is more restrictive than the standard definition of symmetry group of differential equations or finite difference equations where only the solution space is required to be $G$-invariant.  We restrict ourselves to $G$-invariant differential equations to guaranty that these equations can be written in terms of differential invariants so that equation \eqref{invariant pde} holds.
%\end{remark}

\begin{example}\label{heat example - symmetry subgroup}
The heat equation with logarithmic source \eqref{heat} is invariant under the group of transformations
\begin{equation}\label{heat symmetry}
X=x+2\lambda_3e^t+\lambda_2,\qquad T=t+\lambda_1,\qquad \ln U=\ln u-\lambda_3e^tx-\lambda_3^2e^{2t}+\lambda_4e^t,
\end{equation}
where $\lambda_1, \ldots, \lambda_4 \in \mathbb{R}$, \cite{O-1993}.  By direct computation, it is not difficult to see that the numerical scheme \eqref{heat numerical scheme}, on the rectangular mesh \eqref{heat rectangular mesh}, is only invariant under the 3-dimensional group of transformations
$$
X=x+\lambda_2,\qquad T=t+\lambda_1,\qquad \ln U=\ln u +\lambda_4e^t.
$$ 
To see that the one-parameter group 
$$
X=x+2\lambda_3 e^t,\qquad T=t,\qquad \ln U =\ln u -\lambda_3e^tx-\lambda_3^2e^{2t}
$$ 
is not a symmetry, it suffices to note that
\begin{align*}
X_{m,n+1}-X_{m,n}=&2\lambda_3(e^{t_{m,n+1}}-e^{t_{m,n}})+(x_{m,n+1}-x_{m,n})\\
=&2\lambda_3(e^{t_{m,n+1}}-e^{t_{m,n}}) \neq 0\qquad \text{when}\qquad \lambda_3\neq 0.
\end{align*}
\end{example}
As Example \ref{heat example - symmetry subgroup} shows, for a numerical scheme to preserve the symmetry group of  a differential equation, a rectangular mesh might be too restrictive. Consequently, the theory of invariant numerical schemes must be developed over arbitrary meshes. 

%%%%%
\section{Finite Difference Derivatives}\label{discrete derivatives}
%%%%%

In this section we obtain finite difference derivative expressions on an arbitrary mesh.   There are different ways to do so.  One could use the theory of multivariate interpolation \cite{O-2006}, but we prefer to use Taylor polynomial approximations. Depending on the method used, the expressions for the finite difference derivatives can be slightly different, but this difference does not alter the implementation of the moving frame construction discussed in the next section. %(there is an infinite number of discretizations of a given partial derivative).  Similar definitions appear in Dorodnitsyn's book \cite{D-2010} in the form of prolongations of Lie algebras infinitesimal generators. 

To simplify the notation, we assume that the generic point under consideration corresponds to the zero multi-index.  Also, let 
$$
e_i=(0,\ldots,0,1,0,\ldots,0),\qquad i=1,\ldots,p,
$$
be the multi-index of length $p$ with $1$ in the $i^{\text{th}}$ component, and zero elsewhere.  Finally, let 
$$
\Delta_i z_0=z_{0+e_i}-z_0=z_{e_i}-z_0,\qquad i=1,\ldots,p,
$$ 
be the usual forward difference operator in the $i^\text{th}$ component.  

To obtain finite difference approximations for the first order derivatives $u^\alpha_{x^i}$ we consider the first order Taylor polynomial expansions 
\begin{equation}\label{first order Taylor polynomials}
\Delta_i u^\alpha_0 \approx \sum_{j=1}^p \Delta_i x^j_0 \cdot u^\alpha_{x^j} ,\qquad i=1,\ldots,p,\quad \alpha=1,\ldots, q.
\end{equation}
Solving for the first order partial derivatives $u^\alpha_{x^i}$ we obtain
\begin{equation}\label{first order finite difference derivatives}
\begin{pmatrix}
u^\alpha_{x^1} \\ \vdots \\ u^\alpha_{x^p}
\end{pmatrix}
\approx 
\begin{pmatrix}
u^{\alpha,d}_{x^1} \\ \vdots \\ u^{\alpha,d}_{x^p}
\end{pmatrix}
=
\underbrace{\begin{pmatrix}
\Delta_1 x^1_0 & \cdots & \Delta_1x^p_0 \\
\vdots & \ddots & \vdots \\
\Delta_p x^1_0 & \cdots & \Delta_p x^p_0
\end{pmatrix}^{-1}}_{\Delta x^{-1}}
\begin{pmatrix}
\Delta_1 u^\alpha_0\\
\vdots\\
\Delta_p u^\alpha_0
\end{pmatrix},\qquad \alpha=1,\ldots,q,
\end{equation}
which are well defined provided the matrix $\Delta x_0$ is invertible.  When this is the case, the expressions \eqref{first order finite difference derivatives} define first order finite difference derivatives on an arbitrary mesh.  To obtain finite difference approximations for the second order derivatives, it suffices to consider the second order Taylor polynomial expansions
\begin{equation}\label{second order finite difference derivatives1}
\begin{aligned}
\Delta_j \Delta_i u^\alpha_0 &\approx \sum_{k=1}^p \Delta_j\Delta_i x^k_0 \cdot u^\alpha_{x^k}
+\sum_{k,l=1}^p\big[(x^k_{e_i+e_j}-x^k_0)(x^l_{e_i+e_j}-x^l_0)\\
&-\Delta_j x^k_0 \cdot \Delta_j x^l_0 - \Delta_i x^k_0 \cdot \Delta_i x^l_0\big]\frac{u^\alpha_{x^k x^l}}{2},
\end{aligned}
\end{equation}
with $1\leq j \leq i \leq p$ and $\alpha=1,\ldots, q$.  Substituting the first order finite difference derivatives  \eqref{first order finite difference derivatives} in \eqref{second order finite difference derivatives1} and solving for the second order derivatives $u^\alpha_{x^kx^l}$ gives finite difference approximations for second order derivatives.  Those are well defined expressions provided that the matrix with coefficients given by the factors in front of the second order derivatives in \eqref{second order finite difference derivatives1} is invertible.  Continuing in this fashion, it is possible to obtain finite difference approximations for third order derivatives and so on.

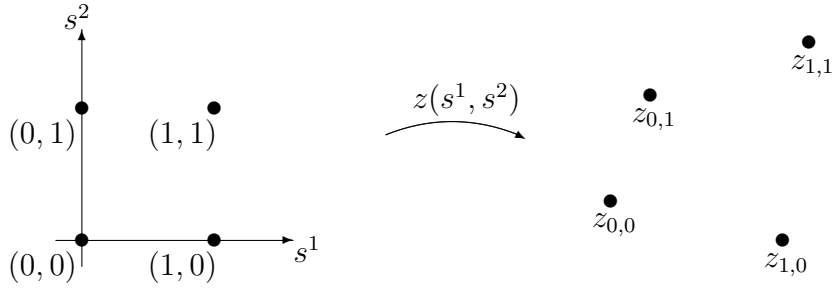
\begin{figure}[!h]
\begin{center}
\begin{picture}(300,90)
%st-frame
\put(0,10){\vector(1,0){90}}
\put(10,0){\vector(0,1){90}}
\put(90,2){$s^1$}
\put(3,90){$s^2$}
\put(10,10){\circle*{5}}
\put(10,60){\circle*{5}}
\put(60,10){\circle*{5}}
\put(60,60){\circle*{5}}
\put(-18,-3){$(0,0)$}
\put(35,-3){$(1,0)$}
\put(-18,47){$(0,1)$}
\put(35,47){$(1,1)$}
\qbezier(125,50)(150,60)(175,50)
\put(173,51){\vector(2,-1){5}}
\put(135,60){$z(s^1,s^2)$}
\put(210,25){\circle*{5}}
\put(203,15){$z_{0,0}$}
\put(275,10){\circle*{5}}
\put(268,0){$z_{1,0}$}
\put(225,65){\circle*{5}}
\put(218,55){$z_{0,1}$}
\put(285,85){\circle*{5}}
\put(278,75){$z_{1,1}$}
\end{picture}
\end{center}
\caption{Multi-index interpretation.}\label{geometric interpretation figure}
\end{figure}

We now consider the continuous limit of those finite difference derivative expressions and verify that they converge to the corresponding partial derivatives.  The standard way of taking the continuous limit of finite difference quantities is to assume that all nodes $z_{N_i}$ coalesce to the same point $z$, and the role of the multi-index $N_i$ is simply to label the points.  In the following, we give a more important role to the multi-index $N_i$.  Instead of viewing the solution of a system of partial differential equations \eqref{pde1} as the graph of a function $(x,u(x))$, we can view a solution as a parametrized $p$-dimensional submanifold
$$
S\to M\simeq X\times U,\qquad s=(s^1,\ldots,s^p)\mapsto z(s)=(x(s), u(s))
$$
transversed to the fibers $\{c\}\times U$.  In this setting, the distinction between the independent variables $x^i$ and the dependent variables $u^\alpha$ disappears.  Moving to the finite difference picture, we now view the multi-indices $N_i$ as forming a square grid with edges of length 1 in the parameter space $S$.  Then, a point $z_{N_i} \in M$ is just the image of $z(s)$ at $s=N_i$, see Figuref \ref{geometric interpretation figure} (in two independent variables).  

In the continuous limit, the point $z_{N_i}=z(N_i)$ converges to $z_0=z(0)=z$ by coalescing the multi-index $N_i$ to the origin.  With this point of view the differences $z_{e_i}-z_0$ converge to $z_{s^i}$ since
$$
\Delta_i z_0 = z_{e_i}-z_0 =\frac{z_{\sigma^i e_i}-z_0}{\sigma^i}\bigg|_{\sigma^i=1} \xrightarrow[]{\sigma^i\to 0} z_{s^i}|_z,\qquad i=1,\ldots, p.
$$
Similarly, in the continuous limit
$$
\Delta_i \Delta_j z_0 \xrightarrow[]{\hskip 0.6cm} z_{s^is^j}|_z,\qquad i,j=1,\ldots, p,
$$
and so on.  It thus follows that the expressions \eqref{first order Taylor polynomials} converge to 
$$
u^\alpha_{s^i}=\sum_{j=1}^p x^j_{s^i} \cdot u^\alpha_{x^j},\qquad \alpha=1,\ldots,q,\quad i=1,\ldots,p,
$$
which is the chain rule formula for $u=u(x(s))$.  Similarly, the expressions \eqref{second order finite difference derivatives1} converge to
\begin{equation}
u^\alpha_{s^i s^j}=\sum_{k=1}^p x_{s^i s^j}^k u^\alpha_{x^k}+\sum_{k,l=1}^p x^k_{s^i} x^l_{s^j} u^\alpha_{x^k x^l},
\end{equation}
in the continuous limit.

\begin{example}
The above discussion is now specialized to the particular situation of two independent variables $(x,y)$ and one dependent variable $u=u(x,y)$.  Without loss of generality, all expressions are centered around $z_{0,0}=(x_{0,0},y_{0,0},u_{0,0})$, and neighboring points are denoted by
$$
z_{m,n}=(x_{m,n},y_{m,n},u_{m,n}),\qquad m,n \in \mathbb{Z}.
$$
Also, let
\begin{equation}\label{shift operator definition}
\Delta z_{m,n}=z_{m+1,n} - z_{m,n},\qquad \delta z_{m,n}=z_{m,n+1} - z_{m,n},
\end{equation}
denote the standard forward difference operators in the two indices and let $(s^1,s^2)=(s,t)$ be coordinates for the parameter space $S$.  

The indices involved in the expressions of the first and second order discrete partial derivatives are displayed in Figure 2. In general, the definition of the $n^\text{th}$ order discrete derivatives will involve the indices contained in the right triangle formed by the origin and the vertices $(n,0)$, $(0,n)$.

\begin{figure}[!h]
\centering
\subfloat[First order derivatives]{\setlength{\unitlength}{0.02cm}
\begin{picture}(200,210)
\put(0,20){\vector(1,0){200}}
\put(10,10){\vector(0,1){200}}
\put(195,25){$s$}
\put(12,205){$t$}
\multiput(10,100)(10,0){8} {\line(1,0){5}}
\multiput(90,20)(0,10){8}{\line(0,1){5}}
\put(10,20){\circle*{5}}
\put(10,100){\circle*{5}}
\put(90,20){\circle*{5}}
\put(90,100){\circle*{5}}
\put(10,180){\circle*{5}}
\put(170,20){\circle*{5}}
\put(-12,0){$(0,0)$}
\put(68,0){$(1,0)$}
\put(-38,95){$(0,1)$}
\put(95,95){$(1,1)$}
\put(148,0){$(2,0)$}
\put(-38,175){$(0,2)$}
{\thicklines
\color{red}\put(10,100){\line(0,-1){80}}
\color{red}\put(10,20){\line(1,0){80}}
\color{red}\put(10,100){\line(1,-1){80}}
\color{red}\put(90,20){\line(-1,1){80}}}
\end{picture}}
\hskip 2cm
\subfloat[Second order derivatives]{\setlength{\unitlength}{0.02cm}
\begin{picture}(200,210)
\put(0,20){\vector(1,0){200}}
\put(10,10){\vector(0,1){200}}
\put(195,25){$s$}
\put(12,205){$t$}
\multiput(10,100)(10,0){8} {\line(1,0){5}}
\multiput(90,20)(0,10){8}{\line(0,1){5}}
\put(10,20){\circle*{5}}
\put(10,100){\circle*{5}}
\put(90,20){\circle*{5}}
\put(90,100){\circle*{5}}
\put(10,180){\circle*{5}}
\put(170,20){\circle*{5}}
\put(-12,0){$(0,0)$}
\put(68,0){$(1,0)$}
\put(-38,95){$(0,1)$}
\put(95,95){$(1,1)$}
\put(148,0){$(2,0)$}
\put(-38,175){$(0,2)$}
{\thicklines
\color{red}\put(90,100){\line(1,-1){80}}
\color{red}\put(90,100){\line(-1,1){80}}
\color{red}\put(10,180){\line(0,-1){160}}
\color{red}\put(10,20){\line(1,0){160}}}
\end{picture}}
\caption{Indices used to define first and second order discrete partial derivatives.}\label{Points for discrete partial derivatives}
\end{figure}
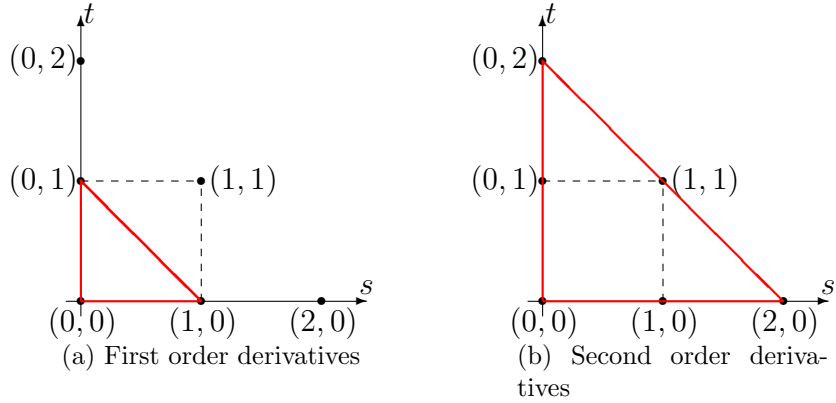

Following our general procedure, to obtain the first order finite difference approximations of $u_x$ and $u_y$ on an arbitrary mesh, the two Taylor expansions 
$$
u_s\approx \Delta u_{0,0} \approx  \Delta x_{0,0} \cdot u_x + \Delta y_{0,0} \cdot  u_y,\qquad
u_t\approx \delta u_{0,0} \approx  \delta x_{0,0} \cdot u_x + \delta y_{0,0} \cdot u_y
$$
are considered.  Solving for $u_x$ and $u_y$ we obtain 
\begin{equation}\label{first order discrete derivatives}
u_x \approx  u_x^d= \frac{\delta y_{0,0} \cdot \Delta u_{0,0} - \Delta y_{0,0} \cdot \delta u_{0,0}}{\Delta x_{0,0} \cdot \delta y_{0,0} - \delta x_{0,0} \cdot \Delta y_{0,0}},\qquad
u_y \approx  u_y^d = \frac{\Delta x_{0,0} \cdot \delta u_{0,0} - \delta x_{0,0} \cdot \Delta u_{0,0}}{\Delta x_{0,0} \cdot \delta y_{0,0} - \delta x_{0,0} \cdot \Delta y_{0,0}}.
\end{equation}
In the continuous limit the expressions \eqref{first order discrete derivatives} converge to
\begin{equation}\label{xy first order derivatives}
u_x = \frac{y_t u_s - y_s u_t}{x_s y_t - x_t y_s},\qquad u_y = \frac{x_s u_t - x_t u_s}{x_s y_t - x_t y_s},
\end{equation}
which are the usual formulas obtained by the chain rule if $x=x(s,t)$, $y=y(s,t)$. 

To obtain approximations for the second order derivatives, it suffices to take second order Taylor expansions:
\begin{align}
u_{ss}\approx \Delta^2 u_{0,0} &\approx \Delta^2x_{0,0} \cdot u_x+\Delta^2y_{0,0}\cdot u_y+[(x_{2,0}-x_{0,0})^2-2(\Delta x_{0,0})^2]\frac{u_{xx}}{2}\nonumber\\
&+[(x_{2,0}-x_{0,0})(y_{2,0}-y_{0,0})-2\Delta x_{0,0} \cdot \Delta y_{0,0}]u_{xy}\nonumber\\
&+[(y_{2,0}-y_{0,0})^2-2(\Delta y_{0,0})^2]\frac{u_{yy}}{2}, \nonumber\\
u_{st}\approx \delta \Delta u_{0,0} &\approx \delta \Delta x_{0,0} \cdot u_x+\delta \Delta y_{0,0} \cdot u_y+[(x_{1,1}-x_{0,0})^2-(\Delta x_{0,0})^2 - (\delta x_{0,0})^2]\frac{u_{xx}}{2}\nonumber\\
&+[(x_{1,1}-x_{0,0})(y_{1,1}-y_{0,0})-\Delta x_{0,0} \cdot \Delta y_{0,0} - \delta x_{0,0} \cdot \delta y_{0,0}]u_{xy}\label{mn discrete second order derivatives}\\
&+[(y_{1,1}-y_{0,0})^2-(\Delta y_{0,0})^2 - (\delta y_{0,0})^2]\frac{u_{yy}}{2},\nonumber\\
u_{tt}\approx \delta^2 u_{0,0} &\approx \delta^2x_{0,0} \cdot u_x+\delta^2y_{0,0} \cdot u_y+[(x_{0,2}-x_{0,0})^2-2(\delta x_{0,0})^2]\frac{u_{xx}}{2}\nonumber\\
&+[(x_{0,2}-x_{0,0})(y_{0,2}-y_{0,0})-2\delta x_{0,0} \cdot \delta y_{0,0}]u_{xy}\nonumber\\
&+[(y_{0,2}-y_{0,0})^2-2(\delta y_{0,0})^2]\frac{u_{yy}}{2}. \nonumber
\end{align}
Solving for the second order derivatives $u_{xx}$, $u_{xy}$, $u_{yy}$ in \eqref{mn discrete second order derivatives}, and replacing the first order derivatives $u_x$ $u_y$ with the approximations \eqref{first order discrete derivatives}, the expressions for the discrete second order derivatives are
\begin{equation}\label{second order finite difference derivatives}
\begin{pmatrix}
u_{xx} \\ u_{xy} \\ u_{yy}
\end{pmatrix}
\approx
\begin{pmatrix}
u_{xx}^d \\ u_{xy}^d \\ u_{yy}^d 
\end{pmatrix}
=
H^{-1} V,
\end{equation}
where $V$ is the column vector
$$
V=
\begin{pmatrix}
\Delta^2 u_{0,0} - \Delta^2 x_{0,0} \cdot u_x^d - \Delta^2 y_{0,0} \cdot u_y^d\\
\delta \Delta u_{0,0}-\delta\Delta x_{0,0}\cdot u_x^d - \delta \Delta y_{0,0} \cdot u_y^d\\
\delta^2 u_{0,0} - \delta^2 x_{0,0} \cdot u_x^d - \delta^2 y_{0,0} \cdot u_y^d
\end{pmatrix},
$$
and $H$ is the $3\times 3$ matrix with entries
\begin{align*}
&H_{11}=[(x_{2,0}-x_{0,0})^2-2(\Delta x_{0,0})^2]/2,& &\hskip -4cm
H_{1,2}=(x_{2,0}-x_{0,0})(y_{2,0}-y_{0,0})-2\Delta x_{0,0} \cdot \Delta y_{0,0},\\
&H_{1,3}=[(y_{2,0}-y_{0,0})^2-2(\Delta y_{0,0})^2]/2,& &\hskip -4cm
H_{2,1}=[(x_{1,1}-x_{0,0})^2-(\Delta x_{0,0})^2-(\delta x_{0,0})^2]/2,\\
&H_{3,3}=[(y_{0,2}-y_{0,0})^2-2(\delta y_{0,0})^2]/2,& &\hskip -4cm
H_{2,3}= [(y_{1,1}-y_{0,0})^2-(\Delta y_{0,0})^2 - (\delta y_{0,0})^2]/2,\\
&H_{3,1}=[(x_{0,2}-x_{0,0})^2-2(\delta x_{0,0})^2]/2,& &\hskip -4cm
H_{3,2}=(x_{0,2}-x_{0,0})(y_{0,2}-y_{0,0})-2\delta x_{0,0} \cdot \delta y_{0,0},\\
&H_{2,2}=(x_{1,1}-x_{0,0})(y_{1,1}-y_{0,0})-\Delta x_{0,0} \cdot \Delta y_{0,0} - \delta x_{0,0} \cdot \delta y_{0,0}.
\end{align*}
Continuing in this fashion, it is possible to obtain higher order finite difference derivatives on an arbitrary mesh.  
\end{example}

The finite difference derivatives constructed above are first order approximations of the continuous derivatives.  More accurate approximations can be obtained by using more nodes to approximate the derivatives.  Also, those approximations are constructed solely with the forward difference operators $\Delta_i$.  To obtain centered difference approximations one can use the right and left differences operators
$$
\Delta_i^{\pm}z_{N_l}=\pm(z_{N_l \pm e_i}-z_{N_l}).
$$
For better numerical schemes, centered difference approximations are used in the next sections.

%%%%%
\section{Equivariant Moving Frames}\label{mf section}  
%%%%%

In this section we review the moving frame construction on joint space.  The exposition follows \cite{O-2001}.  

\begin{definition}
Let $G$ be a finite-dimensional Lie group acting smoothly on a manifold $M$. A \emph{right moving frame} is a smooth $G$-equivariant map $\rho\colon M \to G$ such that 
$$
\rho(g\cdot z)=\rho(z) \cdot g^{-1}\qquad \text{for all}\qquad z \in M,\quad g\in G.
$$
\end{definition}

\begin{theorem}
A right moving frame exists in a neighborhood of a point $z \in M$ if and only if $G$ acts freely and regularly near $z$.
\end{theorem} 

The action is free if at every point $z \in M$ the isotropy subgroup $G_z=\{ e \}$ is trivial.  Under this assumption, the group orbits are of dimension $r=\text{dim }G$.  The action is regular if the orbits form a regular foliation.  The moving frame construction follows from the following theorem.

\begin{theorem}
If $G$ acts freely and regularly on $M$, and $\mathcal{K} \subset M$ is a cross-section to the group orbits, then the right moving frame $\rho\colon M \to G$ at $z \in M$ is defined as the unique group element $g=\rho(z)$ which sends $z$ to the cross-section $\rho(z) \cdot z \in \mathcal{K}$. 
\end{theorem} 

While it is not necessary, we assume $\mathcal{K}=\{ z^1=c^1,\ldots,z^r=c^r \}$ is a coordinate cross-section obtained by fixing the first $r$ coordinates $z=(z^1,\ldots,z^m)$ of $M$ to some suitable constants.  The moving frame $\rho(z)$ is then obtained by solving the \emph{normalization equations}
\begin{equation}\label{normalization equations}
Z^1=g\cdot z^1= c^1,\qquad\ldots\qquad Z^r=g\cdot z^r = c^r,
\end{equation}
for the group parameters $g=(g_1,\ldots, g_r)$ in terms of $z=(z^1,\ldots,z^m)$.  Given a moving frame, there is a systematic way of associating an invariant to a function.

\begin{definition}
Let $\rho$ be a right moving frame, the \emph{invariantization} of a scalar function $F: M \to \mathbb{R}$ is the invariant function
\begin{equation}\label{invariantization map}
I(z)=\iota(F)(z)=F(\rho(z)\cdot z).
\end{equation}
\end{definition}

\begin{proposition}
The invariantization of the coordinate functions $\iota(z^{r+1}),\ldots,\iota(z^m)$ provide a complete set of $m-r$ functionally independent invariants on $M$.
\end{proposition}

Note that by the moving frame construction, the invariantization of the coordinates defining the normalization equations \eqref{normalization equations} are constant, $\iota(z^1)=c^1$, $\ldots$, $\iota(z^r)=c^r$.  Also, if $I(z)$ is an invariant then $\iota(I)=I$.

For most groups of interest, the action on $M$ fails to be free.  There are two common methods for making an (effective) group action free.  In geometry, this is accomplished by prolonging the action to a jet space $\Jn$ of suitably high order.  The cross-section $\mathcal{K}$ is then an $r$-dimensional submanifold of $\Jn$ and the normalization of the group parameters yields a standard moving frame $\rho\n\colon \Jn \to G$.  The invariantization of the jet coordinates $\iota(z\n)=I\n$ gives a complete set of differential invariants \cite{FO-1999}.  The second possibility is to consider the product action of $G$ on a suitably large joint product $M^{\diamond k}\subset M^{\times k}$.  The moving frame construction yields the \emph{product frame} $\rho^{\diamond k}\colon M^{\diamond k} \to G$, and the invariantization of the coordinate functions $z_{N_i}$ gives a complete set of functionally independent joint invariants \cite{O-2001-2}.  

In \cite{O-2001}, it is shown that the right geometrical space to study symmetry of numerical schemes is that of \emph{multi-space}.  The moving frame construction yields what is called a \emph{multi-frame} and the invariantization map \eqref{invariantization map} gives \emph{multi-invariants}.  For the present discussion, it is enough to know that in the continuous limit $\rho^{\diamond k} \to \rho\n$ provided the moving frames are compatible.  To obtain compatible moving frames, instead of working with the standard product action $Z_{N_l}=(X_{N_l}, U_{N_l})=g\cdot z_{N_l}$ on $M^{\diamond k}$ we should consider the joint action on the the finite difference coordinates
$$
x^i_{0},\, x^i_{N_1},\ldots,x^i_{N_k},\qquad  u^\alpha_{0},\qquad u^{d,(n)},
$$
where $u^{d,(n)}$ denotes the collection of finite difference derivatives $u_J^{\alpha,d}$ of order $\leq n$ on an arbitrary mesh computed in Section \ref{discrete derivatives}.   For the discrete invariants $(H_{0},I^{d,(n)})=\iota(x_{0},u^{d,(n)})$ to converge to the differential invariants $(H,I\n)=\iota(x,u\n)$, the cross-section $\mathcal{K}^{\diamond k}$ defining the product frame must, in the continuous limit, converge to a cross-section $\mathcal{K}\n$ for the prolonged action on $\Jn$.  Assuming the action is transitive on the independent variables $x$, this constraint on $\mathcal{K}^{\diamond k}$ implies that we can only normalize the independent variables at one multi-index, say $X_{0}=c$, \cite {O-2001}. 
  
\begin{example}
To illustrate the above discussion we consider the symmetry group of the heat equation with logarithmic source: $u_t=u_{xx}+u\ln u$. The prolonged action induced by \eqref{heat symmetry} is obtained by implementing the chain rule.  The result is
\begin{equation}
\begin{gathered}
(\ln U)_T=(\ln u)_t-\lambda_3e^tx+\lambda_4e^t-2\lambda_3e^t(\ln u)_x\qquad
(\ln U)_X=(\ln u)_x-\lambda_3e^t,\\
(\ln U)_{XX}=(\ln u)_{xx},
\end{gathered}
\end{equation}
and so on.  Choosing the cross-section  
$$
\mathcal{K}^{(1)}=\{ x = t = \ln u = (\ln u)_x =0\}
$$
and solving the normalization equations  $X=T=\ln U=(\ln U)_X=0$ for the group parameters $\lambda_1,$ $\lambda_2,$ $\lambda_3,$ $\lambda_4$ yields the moving frame
\begin{equation}\label{heat moving frame}
\begin{gathered}
\lambda_1=-t,\qquad \lambda_2=-(x+2(\ln u)_x),\qquad \lambda_3=e^{-t}(\ln u)_x,\\
 \lambda_4=e^{-t}(-\ln u+x(\ln u)_x+(\ln u)_x^2).
\end{gathered}
\end{equation}
By construction
$$
\iota(x)=\iota(t)=\iota(u)=\iota(\ln(u)_x)=0,
$$
and the invariantization process yields the differential invariants
\begin{equation}\label{heat invariants}
I=\iota((\ln u)_t)=\ln u_t-\ln u-(\ln u)_x^2,\qquad
J=\iota((\ln u)_{xx})=(\ln u)_{xx},
\end{equation}
and more.  

We now repeat the computations for the discrete case.  In the following, we assume that  
\begin{equation}\label{t assumption}
t_{1,0}-t_{0,0}=0
\end{equation}
as it simplifies the calculations.   The equality \eqref{t assumption} is compatible with the group action \eqref{heat symmetry} as it is an invariant equation of the joint action.   To obtain a product frame compatible with \eqref{heat moving frame} we choose the cross-section
$$
\mathcal{K}^{\diamond 1}=\{ x_{0,0}=t_{0,0}=u_{0,0}=(\ln u)_x^d=0\},
$$
where, due to the assumption \eqref{t assumption}, $(\ln u)_x^d=\cfrac{\ln u_{1,0}-\ln u_{-1,0}}{x_{1,0}-x_{-1,0}}$.  Solving the corresponding normalization equations for the group parameters $\lambda_1,$ $\lambda_2,$ $\lambda_3,$ $\lambda_4$ gives the product frame
\begin{equation}\label{heat group normalizations}
\begin{gathered}
\lambda_1=-t_{0,0},\qquad \lambda_2=-(x_{0,0}+2 (\ln u)^d_x),\qquad \lambda_3=e^{-t_{0,0}}(\ln u)^d_x,\\
\lambda_4=e^{-t_{0,0}}\left[-\ln u_{0,0}+x_{0,0}(\ln u)_x^d+((\ln u)^d_x)^2\right].
\end{gathered}
\end{equation}
The invariantization map is then completely defined and yields, among many others, the discrete invariants
\begin{equation}\label{discrete heat invariants}
\begin{aligned}
&I^d=\iota((\ln u)_t^d)=\frac{\ln u_{01}-e^{\tau}\ln u_{0,0}}{\tau}-\frac{\sigma}{\tau}e^{\tau}(\ln u)_x^d+\frac{e^{\tau}-e^{2\tau}}{\tau}((\ln u)_x^d)^2,\\
&J^d=\iota((\ln u)_{xx}^d)=(\ln u)_{xx}^d,
\end{aligned}
\end{equation}
where $\sigma=x_{0,1}-x_{0,0}$ and $\tau=t_{0,1}-t_{0,0}$, and
 $$
(\ln u)_{xx}^d=\frac{2}{x_{1,0}-x_{-1,0}}\bigg[ \bigg( \frac{\ln u_{1,0}- \ln u_{0,0}}{x_{1,0}-x_{0,0}}\bigg) -  \bigg( \frac{\ln u_{0,0}-\ln u_{-1,0}}{x_{0,0}-x_{-1,0}}\bigg) \bigg].
 $$
The discrete invariant 
\begin{gather}\label{heat mesh}
\iota(x_{0,1})=\sigma+2(e^{\tau}-1)(\ln u)_x^d
\end{gather}
will also prove to be useful in the construction of an invariant numerical scheme in the following section.

\end{example}

%%%%%
\section{Invariant Numerical Schemes}
%%%%%

Let $\Delta_\nu(x,u\n)=0$ be a $G$-invariant system of differential equations as defined in Definition \ref{symmetry definition}.  The invariance of the system guarantees that it can be expressed in terms of the normalized invariants $\iota(x,u\n)=(H,I\n)$.  Using the invariantization map \eqref{invariantization map} 
\begin{equation}\label{invariant pde}
0=\Delta_\nu(x,u\n)=\iota(\Delta_\nu(x,u\n))=\Delta_\nu(H,I\n).
\end{equation}
To obtain an invariant numerical scheme we simply need to replace the differential invariants $(H,I\n)$ in \eqref{invariant pde} by their invariant discrete counterparts $\iota(x_0,u^{d,(n)})=(H_0,I^{d,(n)})$ and add equations specifying the mesh. If the equations determining the mesh are invariant under the group action $G$, the scheme is said to be \emph{fully invariant}, otherwise it is called \emph{partially invariant}.   In applications, invariant constraints on the mesh are usually specified with the aim of simplifying the expressions of the invariant scheme.  This is illustrated in Examples \ref{invariant heat scheme example} and \ref{invariant burgers scheme example}.

An alternative way of constructing the same invariant numerical scheme as above consists of replacing the partial derivatives in the differential equations $\Delta_\nu(x,u\n)=0$ by their finite difference approximation
\begin{equation}\label{num scheme}
\Delta_\nu(x,u\n)=0 \qquad \longrightarrow \qquad E_\nu=\Delta_\nu(x_{0,0},u^{d,(n)})=0,
\end{equation}
followed by the invariantization of \eqref{num scheme}. 

%By construction of the discrete derivatives in Section \ref{discrete derivatives}, the invariant scheme obtained is well defined on any compatible mesh.

\begin{example}\label{invariant heat scheme example}
The heat equation with source \eqref{heat} is easily expressed in terms of the differential invariants \eqref{heat invariants}.  By invariantizing \eqref{heat}, the equation can be rewritten as
\begin{equation}\label{invariant heat}
0=\iota(u_t - u_{xx} - u\ln u)=I-J.
\end{equation}

To obtain an invariant discrete version, $I$ and $J$ in \eqref{invariant heat} are simply replaced by their discrete counterparts: 
\begin{equation}\label{discrete heat general mesh}
I^d-J^d=\frac{\ln u_{0,1}-e^{\tau}\ln u_{0,0}}{\tau}-\frac{\sigma}{\tau}e^{\tau}(\ln u)_x^d+\frac{e^{\tau}-e^{2\tau}}{\tau}((\ln u)_x^d)^2-(\ln u)_{xx}^d=0.
\end{equation}
By construction, the numerical scheme is a valid approximation of \eqref{heat} on any mesh satisfying the flat time assumption \eqref{t assumption}.  At the moment, there is no systematic method for obtaining invariant equations defining the mesh of an invariant numerical scheme.  In \cite{K-2008} it is claimed that the equations for the invariant mesh are simply obtained by invariantizating the equations of the mesh equations for the non-invariant numerical scheme.  The issue with this proposition is that there are no guarantee that the equations obtained are compatible.   Indeed, this idea was used in \cite{CH-2011} and led to the incompatible mesh equations (83--85).  To illustrate the problem we invariantize the equations \eqref{heat rectangular mesh} describing a uniform rectangular mesh.  The result is
\begin{align*}
&x_{1,0}-x_{0,0}=h,& & x_{0,1}-x_{0,0}=\sigma=2(\ln u)_x^d(1-e^\tau),\\ 
&t_{1,0}-t_{0,0}=0,& &t_{0,1}-t_{0,0}=\tau=k.
\end{align*}
In the time variable $t$ the mesh equations are compatible but this is not the case in the spatial variable $x$.  By shifting the first equation by $\delta$ and the second equation by $\Delta$ (recall \eqref{shift operator definition}) we obtain the constraint
\begin{equation}\label{constraint}
h=x_{1,1}-x_{0,1}=2 \Delta[(\ln u)_x](1-e^k).
\end{equation}
Since $h$ and $k$ are constants, the equation \eqref{constraint} imposes a restriction on the solution $u$ which is not admissible.  To circumvent the problem we can neglect the equation $x_{10}-x_{0,0}=h$ and assume that the mesh equations are given by
\begin{equation}\label{discrete heat particular mesh}
\sigma=2(1-e^{\tau})(\ln u)_x^d,\qquad \tau=k,\qquad \Delta t=0.
\end{equation}
With \eqref{discrete heat particular mesh} the mesh is uniquely determined once the sample points in $x$ are fixed at some time $t=t_0$.  %As time evolves the sample points in $x$ evolve according to the first equation in \eqref{discrete heat particular mesh}.  
In conclusion, a fully invariant numerical numerical scheme for the heat equation \eqref{heat} is given by
\begin{equation}\label{discrete heat}
\begin{gathered}
E_1=\frac{\ln u_{0,1}-e^{\tau}\ln u_{0,0}}{\tau}-e^{\tau}\frac{1-e^{\tau}}{\tau}((\ln u)_x^d)^2-(\ln u)_{xx}^d=0,\\
E_2=\sigma-2(1-e^{\tau})(\ln u)_x^d=0,\qquad  E_3=\tau - k = 0, \qquad E_4=\Delta t =0.
\end{gathered}
\end{equation}
\end{example}

\begin{example}\label{invariant burgers scheme example}
In this second example we find an invariant numerical scheme for the spherical Burgers equation
\begin{equation}\label{monge-ampere}
u_t+\frac{u}{t}+uu_x+u_{xx}=0, \qquad t>0.
\end{equation}
The equation \eqref{monge-ampere} admits the three-dimensional Lie algebra of infinitesimal symmetry generators, \cite{I-1993}, 
$$
\vv_1=\pp{}{x},\qquad \vv_2=x\pp{}{x}+2t\pp{}{t}-u\pp{}{u},\qquad \vv_3=\ln t\pp{}{x}+\frac{1}{t}\pp{}{u}.
$$
The corresponding group action is
\begin{equation}\label{ma action}
X=e^{\lambda_2}(x+\lambda_3\ln t) + \lambda_1,\qquad
T=e^{2\lambda_2} t,\qquad
U= e^{-\lambda_2} \left(u+\frac{\lambda_3}{t}\right).
\end{equation}
For the joint action, a simple choice of cross-section is 
$$
x_{0,0}=0,\qquad t_{0,0}=1,\qquad u_{0,0}=0.
$$
Solving the normalization equations
$$
X_{0,0}=0,\qquad T_{0,0}=1,\qquad U_{0,0}=0
$$
for the group parameters yields the product frame
\begin{equation}\label{ma multi-frame}
\lambda_1=-\frac{x_{0,0}-u_{0,0}t_{0,0}\ln t_{0,0}}{\sqrt{t_{0,0}}},\qquad
\lambda_2=\ln\left(\frac{1}{\sqrt{t_{0,0}}}\right), \qquad
\lambda_3=-u_{0,0}t_{0,0}. 
\end{equation}

An invariant numerical scheme for the spherical Burgers equation \eqref{monge-ampere} is given by the invariantization of the discrete derivatives $u^d_{xx}$ and $u^d_t$:
\begin{gather}\label{discrete invariant Burgers}
0=I^d=\iota(u^d_{t})+\iota(u^d_{xx})=u_t^d+\frac{u_{0,0}}{t_{0,0}}+u_{0,0}u_x^d+u_{xx}^d.
\end{gather}
The expression \eqref{discrete invariant Burgers} is defined on any compatible mesh.  We now impose some invariant constraint on the mesh which will simplify the coordinate expressions of \eqref{discrete invariant Burgers}.  Once more, we assume that $\Delta t_{0,0}=t_{1,0}-t_{0,0}=0$ as this constraint is invariant under the group action \eqref{ma action}.  Another invariant constraint is given by $\delta^2 t_{0,0}= t_{0,2}-2t_{0,1}+t_{0,0}=0$.  Finally, using the invariant
$$
\iota(x_{0,1})=x_{0,1}-x_{0,0}-ut\ln\bigg(\frac{t_{0,1}}{t_{0,0}}\bigg)=\sigma-u_{0,0} t_{0,0} \ln \bigg(\frac{t_{0,1}}{t_{0,0}}\bigg),
$$
an invariant mesh is specified by the equations
\begin{equation}\label{discrete burger particular mesh}
\sigma=u_{0,0}t_{0,0}\ln\bigg(\frac{t_{0,1}}{t_{0,0}}\bigg),\qquad \Delta t_{0,0}=0,\qquad \delta^2 t_{0,0} =0.
\end{equation}
As in the previous example, once the sample points in $x$ are determined at some time $t=t_0$ and the step size in $\tau=k$ is chosen, the equations \eqref{discrete burger particular mesh} uniquely fix the mesh.   On the mesh \eqref{discrete burger particular mesh} the equation \eqref{discrete invariant Burgers} simplifies to the fully invariant numerical scheme
\begin{equation}\label{discrete Burgers}
\begin{gathered}
E_1=u_{0,1}-\frac{u_{0,0}t_{0,0}}{t_{0,1}}+2\tau u_{xx}^d=0,\\
E_2=\sigma-u_{0,0}t_{0,0}\ln\bigg(\frac{t_{0,1}}{t_{0,0}}\bigg)=0,\qquad  E_3=\Delta t_{0,0} = 0, \qquad E_4=\delta^2 t_{0,0} =0,
\end{gathered}
\end{equation}
where
$$
u_{xx}^d=\frac{2}{x_{1,0}-x_{-1,0}}\bigg[ \bigg( \frac{u_{1,0}-u_{0,0}}{x_{1,0}-x_{0,0}}\bigg) -  \bigg( \frac{u_{0,0}-u_{-1,0}}{x_{0,0}-x_{-1,0}}\bigg) \bigg].
$$
\end{example}

\section{Numerical Tests}

Fully invariant numerical schemes for the heat equation with logarithmic source \eqref{heat} and the spherical Burgers equation \eqref{monge-ampere} were obtained in \eqref{discrete heat} and \eqref{discrete Burgers}, respectively.  To illustrate how preservation of symmetry can increase the accuracy of a numerical method, we compare the fully invariant schemes with two closely related schemes.  Firstly, recall that provided the flat time assumption \eqref{t assumption} holds, the discretizations \eqref{discrete heat general mesh} and \eqref{discrete invariant Burgers} are valid approximations of the corresponding partial differential equation on any compatible mesh.  To gauge the effect of imposing invariant constraints on the mesh, we compare the fully invariant schemes to partially invariant schemes obtained by restricting \eqref{discrete heat general mesh} and \eqref{discrete invariant Burgers} to non-invariant uniform rectangular meshes.  Then, for further comparison, the fully and partially invariant schemes are set against standard finite difference approximations on uniform rectangular meshes.  

%%%%
\subsection{Heat Equation}
%%%%

\begin{figure}[ht]
\begin{minipage}[b]{0.5\linewidth}
\centering
\includegraphics[width=7cm,height=6cm]{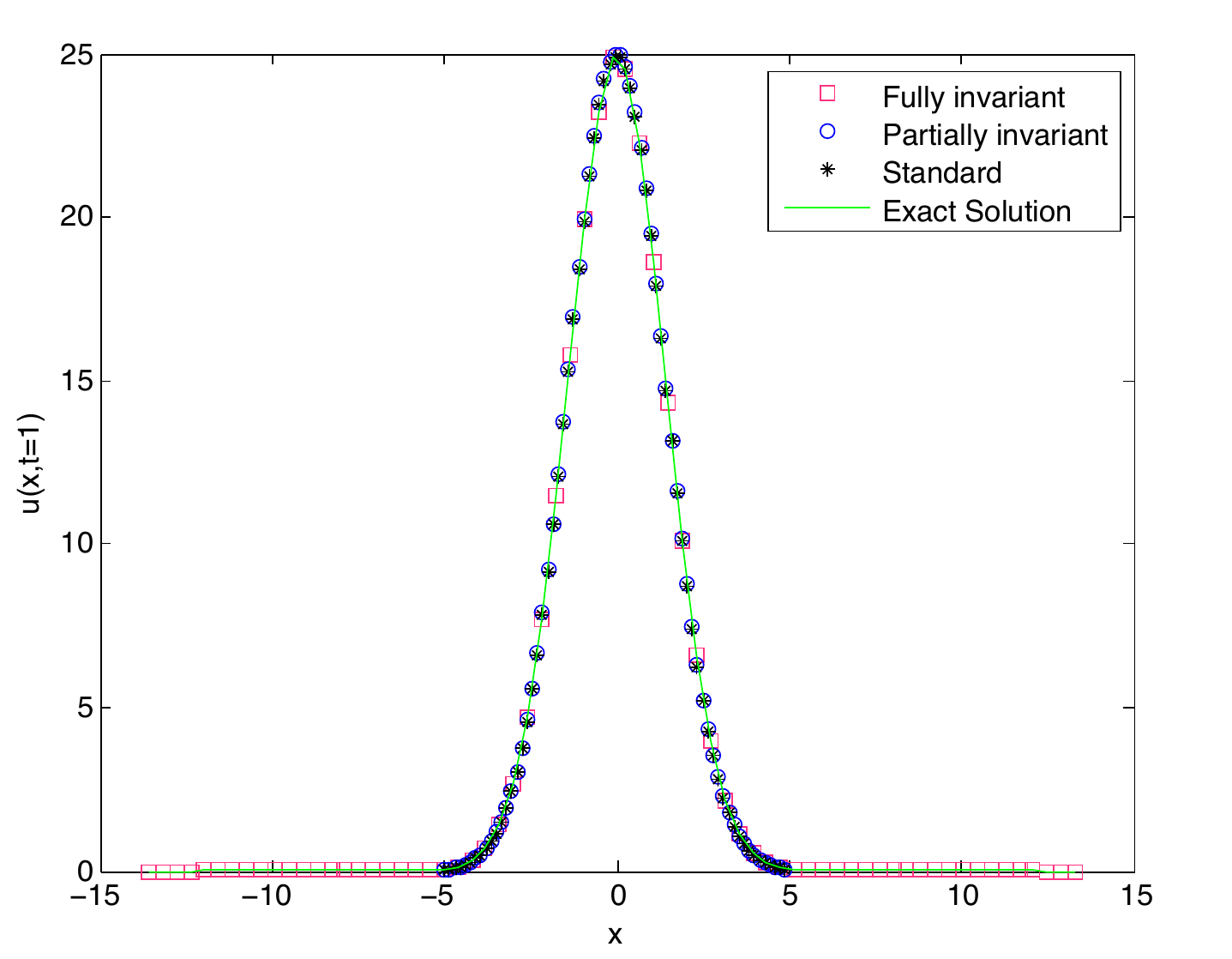}
\caption{Solution $u=\exp[x^2/4+1/2]$.}
\label{figure heatsol}
\end{minipage}
\hspace{0.5cm}
\begin{minipage}[b]{0.5\linewidth}
\centering
\includegraphics[width=7cm,height=6cm]{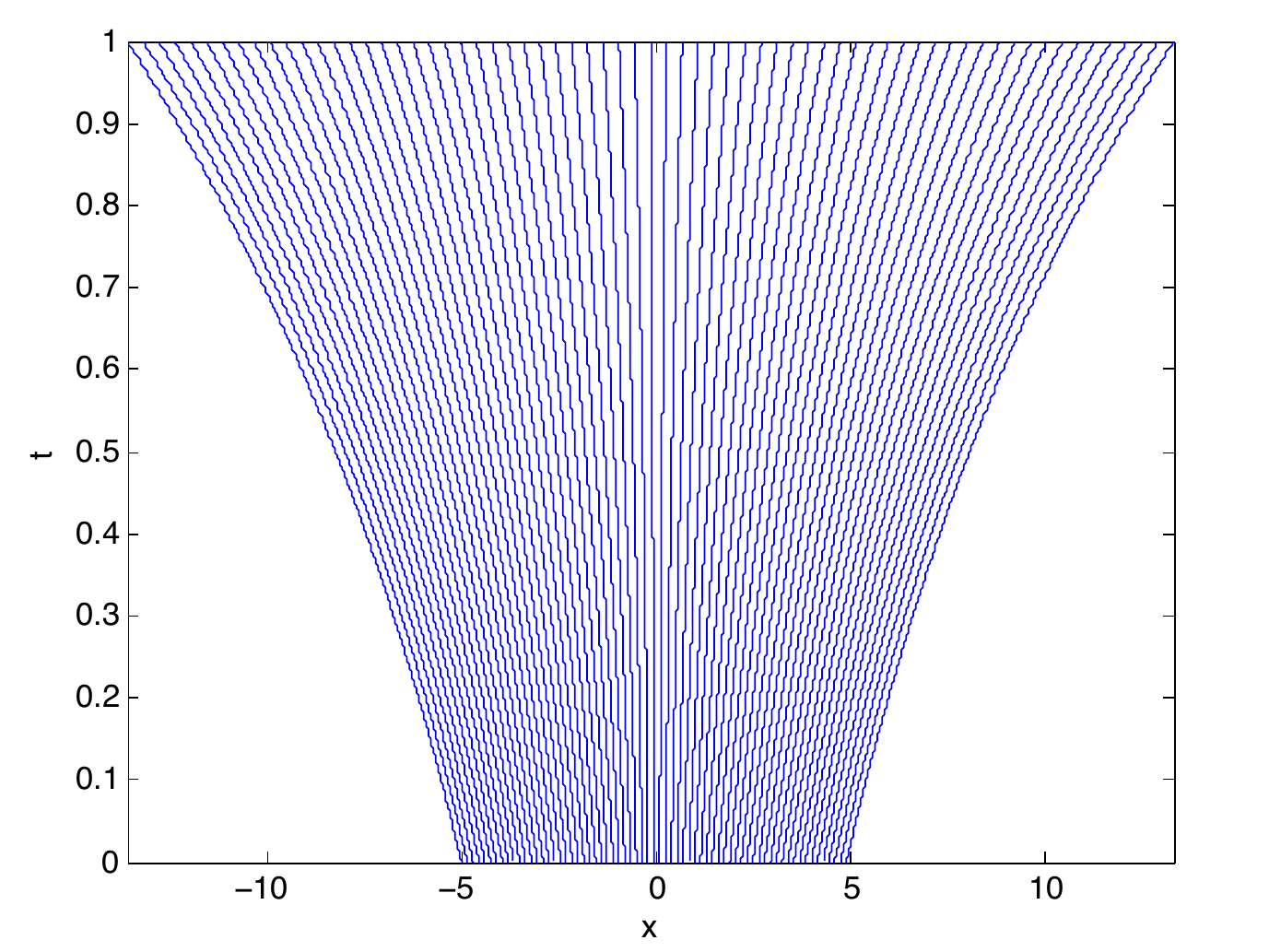}
\caption{Invariant mesh.}
\label{figure heatmesh}
\end{minipage}
\end{figure}

\begin{figure}[ht]
\centering
\includegraphics[width=7cm,height=6cm]{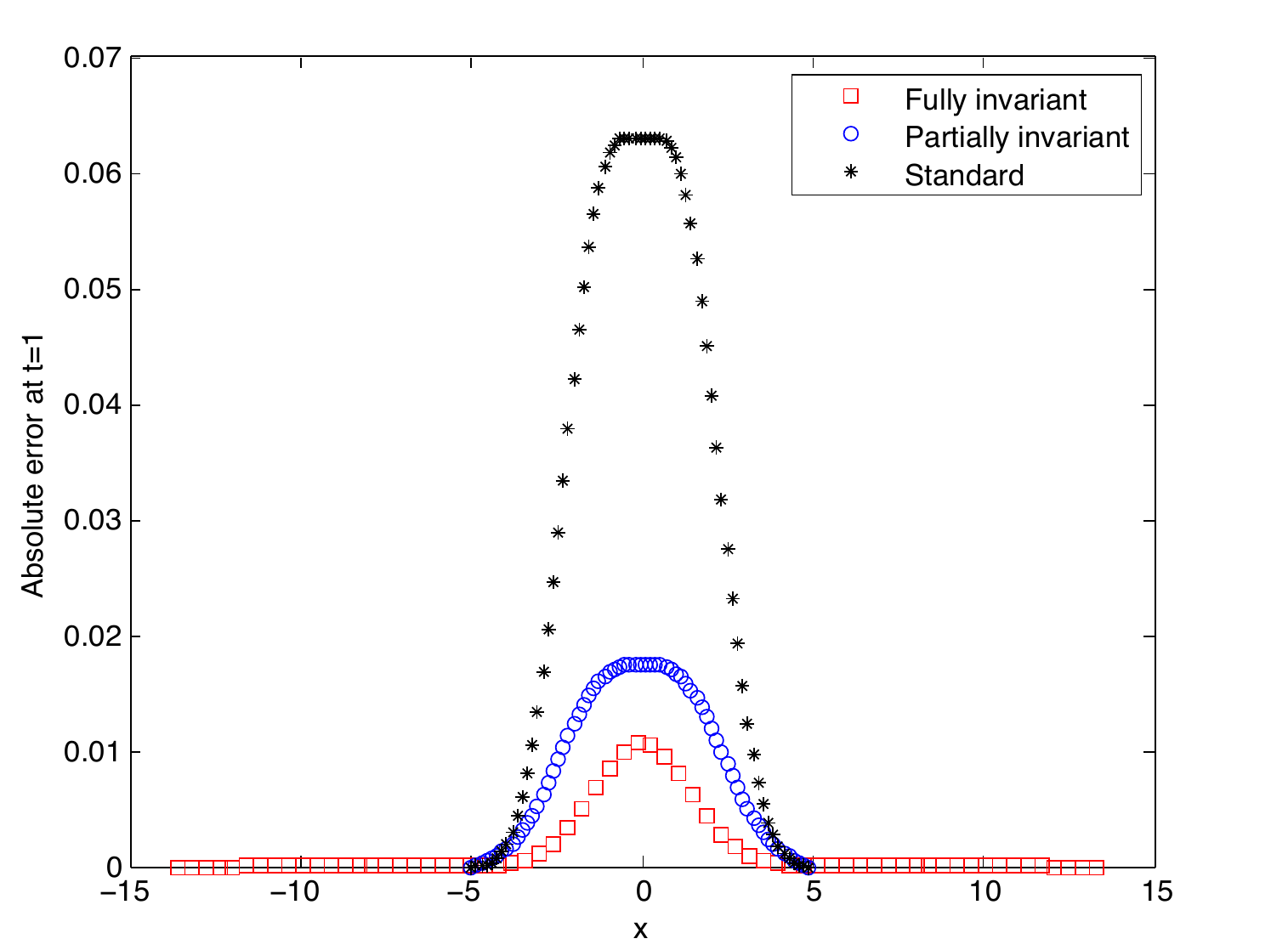}
\caption{Absolute errors for the solution $u=\exp[x^2/4+1/2]$.}
\label{figure heaterror}
\end{figure}

For the heat equation \eqref{heat} we used the steady-state solution
\begin{equation}\label{heat solution test}
u(x,t)=\exp\bigg[ce^t-\frac{x^2}{4}+\frac{1}{2}\bigg]\qquad \text{with}\qquad c=0,
\end{equation}
to compare the precision of the three schemes.  Starting at $t=0$ and working on the space interval $[-5,5]$, the solution after one unit of time elapsed is shown in Figure \ref{figure heatsol}.  The numerical simulation was done using an initial spacial step size of $h=0.15$ and a constant time step $\tau=k=0.001$.  We note that for the fully invariant scheme \eqref{discrete heat} the mesh evolves according to the equation
$$
\sigma=2(1-e^{\tau})(\ln u)_x^d.
$$
The result is an expansion of the space interval $[-5,5]$ over the course of the simulation as shown in Figure \ref{figure heatmesh}.  
The absolute errors for the three numerical schemes are given in Figure \ref{figure heaterror}.  As anticipated, the standard numerical scheme offers the worst precision among the three.   The partially invariant scheme on a rectangular mesh gives better results, but the error decreases even more for the fully invariant numerical scheme.

%%%%%
\subsection{Burgers Equation}
%%%%%

\begin{figure}[ht!]
\begin{minipage}[b]{0.5\linewidth}
\centering
\includegraphics[width=7cm,height=6cm]{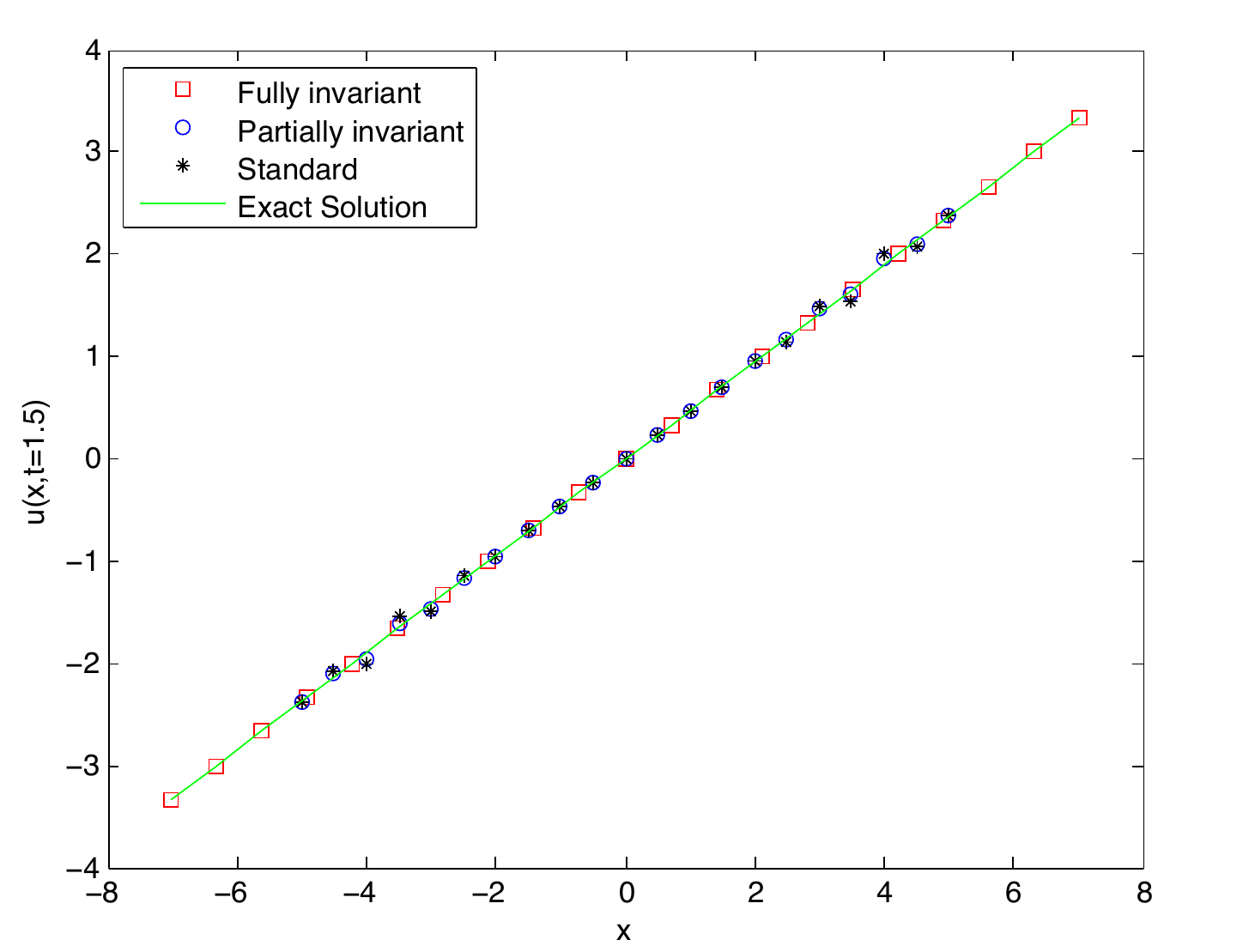}
\caption{Solution $u=x/[t(1+\ln t )]$.}
\label{figure burgerssol}
\end{minipage}
\hspace{0.5cm}
\begin{minipage}[b]{0.5\linewidth}
\centering
\includegraphics[width=7cm,height=6cm]{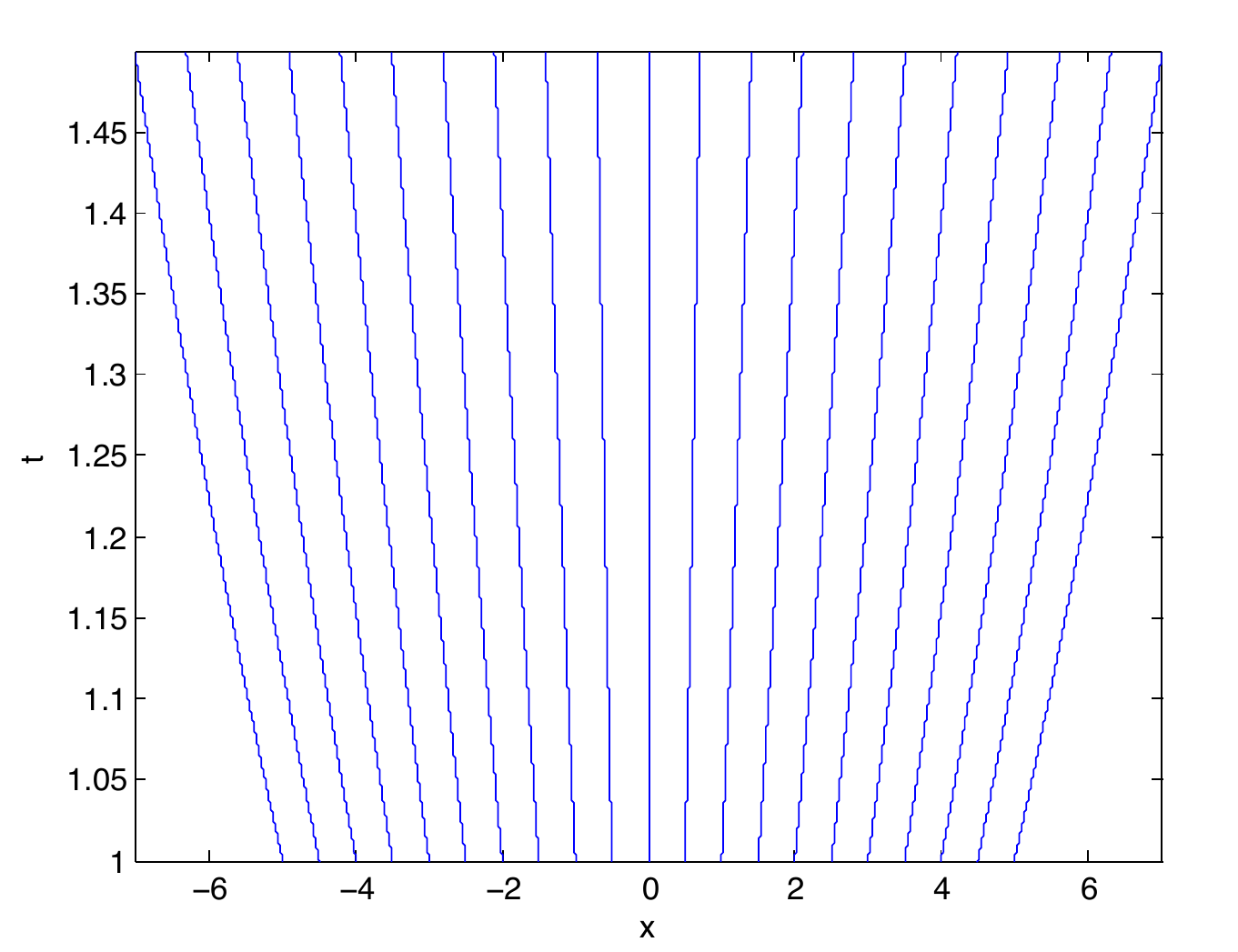}
\caption{Invariant mesh.}
\label{figure burgersmesh}
\end{minipage}
\end{figure}

For the spherical Burgers equation \eqref{monge-ampere}, a numerical test was performed using the solution
$$
u(x,t)=\frac{x+c_1}{t(c_2+\ln t )}\qquad  \text{with}\qquad c_1=0\quad \text{and}\quad c_2=1.
$$
%simulation yields qualitative results similar to those obtained for the heat equation \eqref{heat}.  The numerical 
%$$ 
At $t=1$, the initial step sizes chosen  are $h=0.5$ in $x$ and $\tau=k=0.001$ in $t$.  In Figure \ref{figure burgerssol}, the solution is shown at $t=1.5$.  As in the previous simulation, the mesh of the fully invariant numerical scheme evolves as a function of time.  The evolution is governed by the equation 
$$
\sigma=u_{0,0}t_{0,0}\ln\bigg(\frac{t_{0,1}}{t_{0,0}}\bigg).
$$
The evolutive mesh appears in Figure \ref{figure burgersmesh}.  

\begin{figure}[ht]
\begin{minipage}[b]{0.5\linewidth}
\centering
\includegraphics[width=7cm,height=6cm]{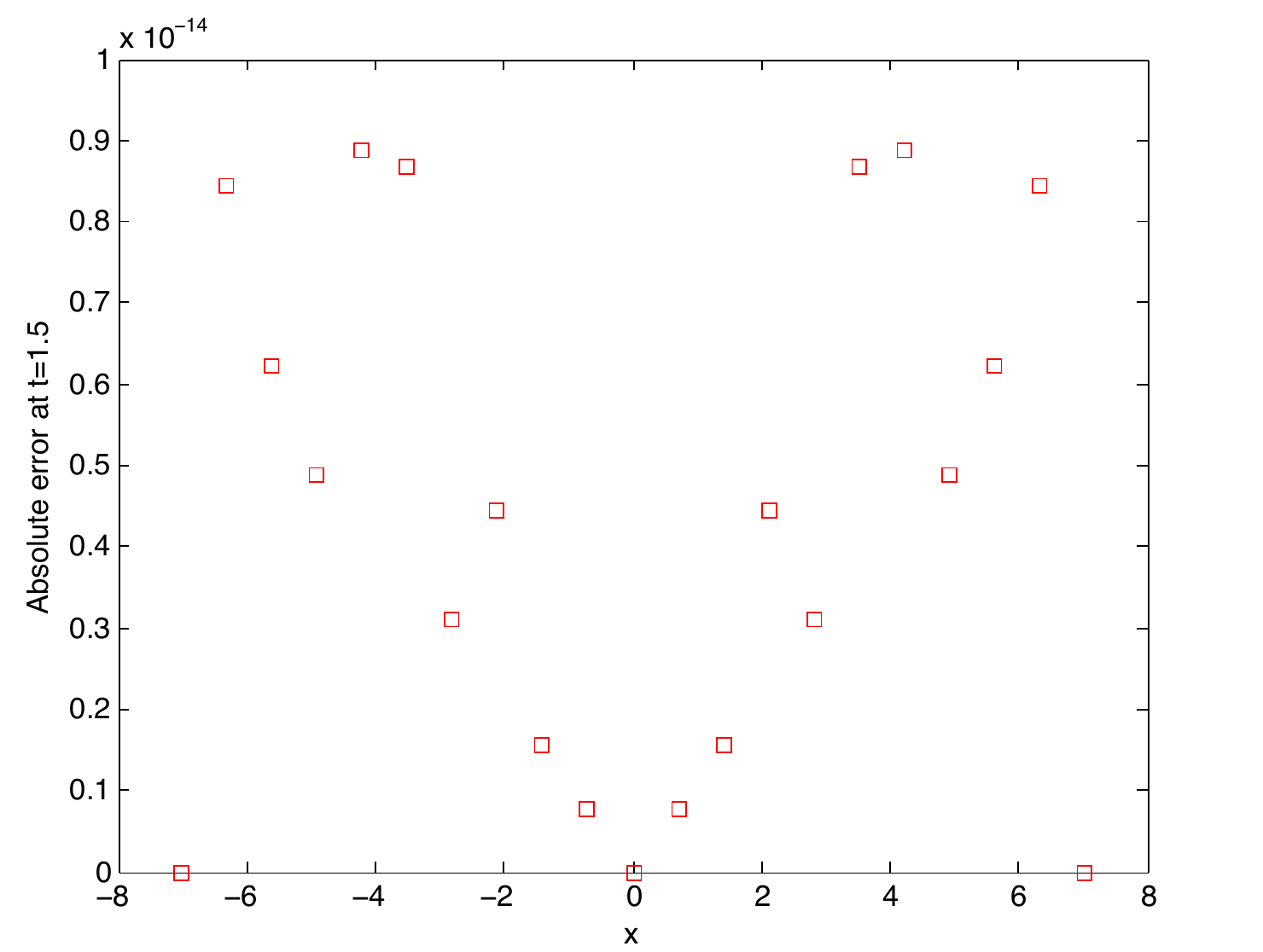}
\caption{Absolute error for the fully invariant scheme.}
\label{figure burgerserror1}
\end{minipage}
\hspace{0.5cm}
\begin{minipage}[b]{0.5\linewidth}
\centering
\includegraphics[width=7cm,height=6cm]{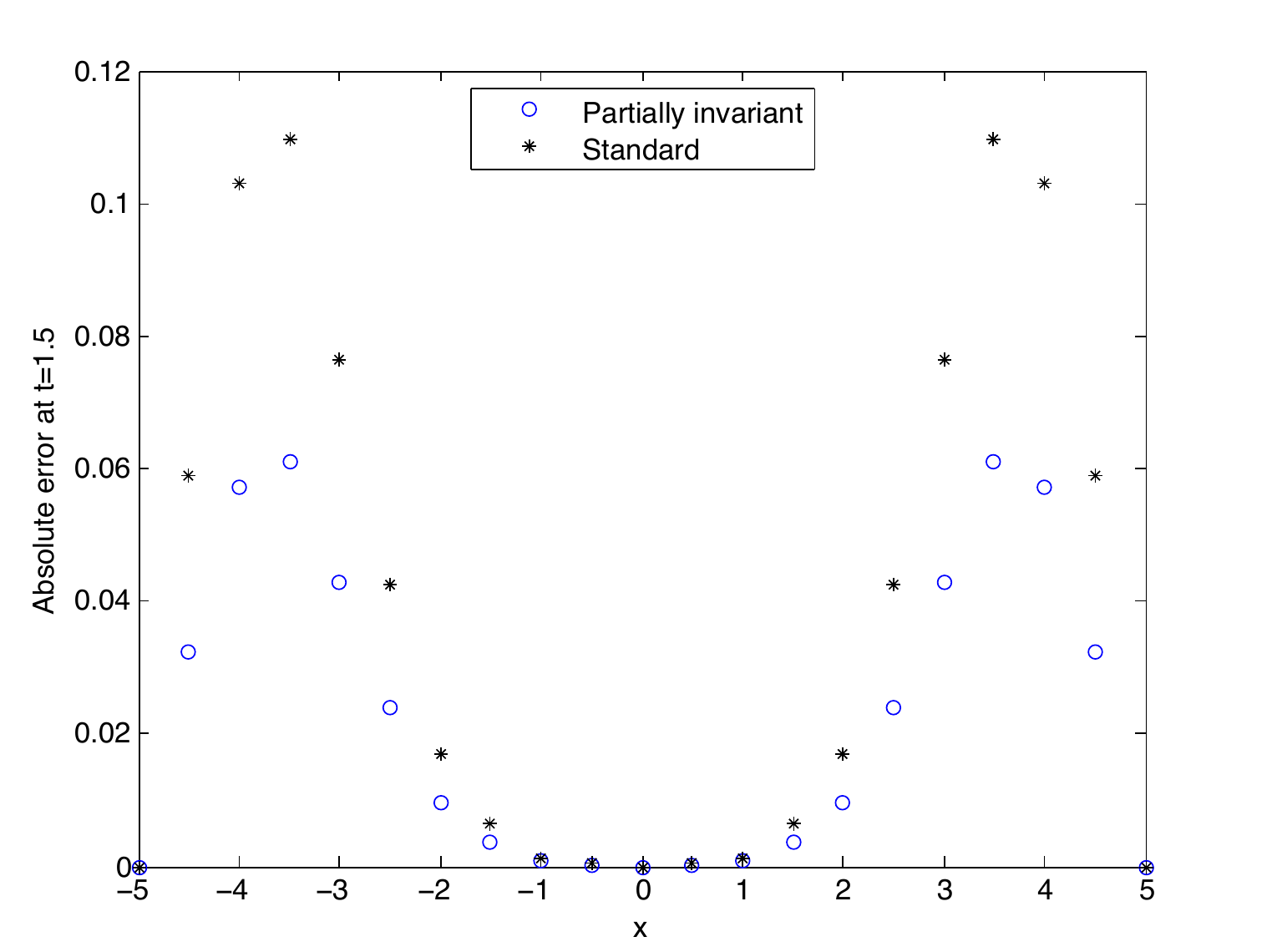}
\caption{Absolute errors for the standard and partially invariant schemes.}
\label{figure burgerserror2}
\end{minipage}
\end{figure}

The absolute errors for the three numerical schemes are given in Figures \ref{figure burgerserror1} and \ref{figure burgerserror2}.  Once more, the fully invariant scheme is more precise.  For the solution considered the improvement is considerable as the error is reduced by a factor $10^{ -13}$ compared with the two other schemes.   On the other hand, while the partially invariant scheme is slightly more accurate than the standard scheme, the errors are comparable.

\section{Concluding Remarks}

Other simulations, not included in the paper, indicate that invariant schemes do not always improve the numerical accuracy of a standard scheme.  While the error is generally not worst than the standard scheme it is still not clear for which types of equations or solutions an invariant scheme will give significantly better results.  Knowing when this is the case would be useful for future applications.

As with all other works on the subject, the important question of the stability of an invariant scheme was not addressed in this paper.  At the moment it is not clear how the evolution of the mesh affects the stability.  %Does the scheme become more or less stable as the mesh evolve? Or does it have no influence?

Finally, many partial differential equations admit infinite-dimensional symmetry groups.  Classical examples include the Kadomtsev--Petviashvili equation \cite{DKLW-1985}, the Infeld--Rowlands equation \cite{FW-1993} and the Davey--Stewartson equations \cite{CW-1988}.  %Plenty of other examples appear in hydrodynamic.  
An interesting problem consists of finding a procedure for obtaining invariant discretizations of partial differential equations admitting infinite-dimensional symmetry groups.

%%%%%
\section*{Acknowledgement}
%%%%%

The authors would like to thank Pavel Winternitz and Decio Levi for their comments and suggestions on the project.

%%%%%

\end{document}